\documentclass[reprint,amsmath,amssymb,aps]{revtex4-2}

\usepackage{graphicx}
\usepackage{dcolumn}
\usepackage{bm}
\usepackage{braket}
\usepackage{booktabs}
\usepackage{multirow}
\usepackage{xcolor}
\usepackage{natbib}
\usepackage{booktabs}

\begin{document}

\title{First application of the transcorrelated method to noncovalent interactions: The A24 dataset}

\author{
Johannes Hauskrecht$^{1}$,
Stephanie Lambie$^{1}$,
Evelin Martine Corvid Christlmaier$^{1}$,
Thomas Schraivogel$^{3}$,
Daniel Kats$^{1}$,
and Ali Alavi$^{1,2}$\\[0.5em]
\small $^{1}$Max Planck Institute for Solid State Research, Stuttgart, Germany\\
\small $^{2}$Yusuf Hamied Department of Chemistry, University of Cambridge,
Lensfield Road, Cambridge CB2 1EW, UK\\
\small $^{3}$Scientific Computing Center, Karlsruhe Institute of Technology,
Karlsruhe, Germany
}

\date{\today}

\begin{abstract}
We present the first application of transcorrelated (TC) coupled-cluster (CC) theory to noncovalent interactions within the xTC approximation. 
The method is assessed for the A24 dataset of hydrogen-bonded, mixed and pure dispersion bound dimers. The xTC interaction  energies are computed at the CCSD, DCSD, and CCSD(T) levels in aug-cc-pVDZ (AVDZ), aug-cc-pVTZ (AVTZ) basis sets, and are compared with both canonical and explicitly correlated F12 methods. 
Because non-covalent interaction energies rely on delicate error cancellation between dimers and monomers, we optimize the TC Jastrow factor for each dimer, and reuse the same parameters for the corresponding monomer calculations. This shared-parameter strategy reduces stochastic optimization noise which would otherwise dominate the interaction energies. The results show that the xTC-CCSD(T)/AVTZ method performs extremely well for the hydrogen-bonded systems, with a mean-absolute error of only 0.007 kcal/mol in the interaction energies, with respect to the benchmark values generated with CCSD(T)/CBS + $\Delta$CCSDT(Q) + core corrections. 
For pure dispersion bound systems the errors are slightly larger (0.055 kcal/mol), leading to an overall MAE of 0.030 kcal/mol for the entire dataset. 
Adding a $\Delta$MP2 correction to the xTC-CCSD(T)/AVDZ brings these results close to AVTZ quality and provides a practical route toward larger noncovalent systems. 
A decomposition of the xTC interaction energy into a mean-field and correlation contribution shows that part of the correlation contribution is systematically shifted by the TC method to the mean-field contribution. This physically appealing feature indicates that the TC method has great potential for the quantitative description of noncovalent interactions, and opens a new route for high-accuracy quantum chemistry to be applied to systems of biological and soft-matter interest. 

\end{abstract}

\maketitle

\section{Introduction}

Noncovalent interactions are central to molecular recognition, solvation, supramolecular chemistry, biomolecular structure, and molecular materials. Their accurate theoretical description is challenging because the interaction energy is obtained as a small difference between large monomer and dimer total energies. Consequently, reliable predictions require not only accurate absolute energies, but also a balanced cancellation of systematic errors between the complex and its fragments. The physical origin of noncovalent binding is also diverse: electrostatics, exchange repulsion, induction, and dispersion contribute with different relative importance in hydrogen-bonded, mixed electrostatic/dispersion, and dispersion-dominated systems. This makes noncovalent interactions a sensitive test of electronic-structure methods. Coupled-cluster theory, in particular CCSD(T), extrapolated to the complete-basis-set limit, is widely regarded as the standard high-accuracy reference approach for small and medium-sized noncovalent complexes \cite{cizek1966coupled,bartlett2007coupled,helgaker2000molecular,rezac2013goldstandard,rezac2016review}. However, for benchmark-level data, where errors below $0.1$ kcal mol$^{-1}$ are desired, even CCSD(T)/CBS may not be sufficient, and corrections from higher-order excitations, core correlation, and relativistic effects can become relevant on the scale of the target accuracy \cite{rezac2013goldstandard,rezac2015extensions}.

A major practical obstacle in such calculations is the slow convergence of the correlation energy with respect to the one-particle basis set. This problem is particularly severe for noncovalent interactions, where augmented basis sets are often required to describe the interfragment region and long-range correlation effects reliably \cite{marshall2011basis,rezac2016review}. Explicitly correlated R12/F12 methods address this bottleneck by introducing the interelectronic distance explicitly into the wavefunction ansatz, thereby accelerating the convergence of short-range pair correlation with the orbital basis \cite{kong2012f12,adler2007ccsdtf12,knizia2009f12}. F12 methods have, therefore, become a standard tool in high-accuracy noncovalent benchmark calculations and provide a natural reference point for any alternative explicitly correlated approach \cite{marshall2011basis,rezac2015extensions}.


The transcorrelated (TC) method offers a different route to explicit correlation. Instead of augmenting the many-body wavefunction directly, a Jastrow factor is used to perform a similarity transformation of the electronic Hamiltonian, producing an effective non-Hermitian Hamiltonian that incorporates part of the electron correlation problem \cite{boys1969tc,handy1969tc}. In an exact treatment, this transformation leaves the spectrum unchanged. In a finite basis and with a truncated many-body solver, however, the transformed Hamiltonian can both accelerate basis-set convergence and modify the effective many-body problem seen by the approximate method.

This second aspect distinguishes transcorrelation from standard F12 corrections. While F12 methods primarily improve the basis-set convergence of a chosen many-body approximation, TC can also compactify the wavefunction and thereby change the behavior of the approximate post-Hartree--Fock method itself. This makes TC particularly interesting for approximate correlation methods, where the Hamiltonian transformation may improve not only the basis-set convergence but also the quality of the truncated many-body expansion.

In recent years, TC methods have been combined with FCIQMC, periodic coupled-cluster theory, density-matrix renormalization group, and coupled-cluster methods, and the xTC approximation has made the treatment of the three-body terms generated by the transformation more practical \cite{luo2018fciqmc,liao2021periodic,baiardi2020tcdmrg,liao2023tcdmrg,schraivogel2021tccc,schraivogel2023tccc2,christlmaier2023xtc}. Related work has also investigated Jastrow optimization and wavefunction compactification strategies for TC Hamiltonians \cite{haupt2023optimizing,ammar2024compactification}.

Despite these developments, the performance of transcorrelated methods for benchmark noncovalent interaction energies has, to the best of our knowledge, not yet been systematically assessed. This is not a trivial extension of existing molecular applications. Noncovalent interaction energies depend on subtle dimer--monomer error cancellation, and the use of an optimized Jastrow factor introduces additional questions of transferability and balance between the complex and its fragments. Moreover, dispersion-dominated systems probe long-range interfragment pair correlation, which may not be described optimally by a short- or medium-range Jastrow parametrization. This means that it is not obvious a priori whether a transcorrelated Hamiltonian that improves total energies or atomization energies will also provide a balanced description of noncovalent binding.

The present work addresses this question for the A24 dataset. Several benchmark sets for noncovalent interactions exist, including S22, S66, L7, and S12L \cite{jurecka2006s22,rezac2011s66,rezac2011s66x8,sedlak2013l7,grimme2012s12l}. We focus here on A24 because it contains small dimers with highly accurate CCSD(T)/CBS-based reference interaction energies, supplemented by higher-level corrections, and spans hydrogen-bonded systems (1--5), mixed electrostatic/dispersion systems (6--15), and dispersion-dominated systems (16--24) \cite{rezac2013goldstandard,rezac2015extensions}. This makes it well suited as a first stringent test of TC for noncovalent interactions: the systems are small enough for systematic coupled-cluster comparisons, but diverse enough to reveal interaction-class-dependent trends.

We compare conventional canonical coupled-cluster calculations, hereafter referred to as plain calculations, with explicitly correlated F12 and reference-corrected xTC (RC-xTC) calculations. The comparison is carried out at the CCSD, distinguishable-cluster singles and doubles (DCSD),\cite{kats2013dca,kats2014} and CCSD(T) levels. The aim is to separate basis-set acceleration from method-level improvement. In particular, we ask whether xTC can compete with F12 for A24 interaction energies, whether its effect depends on the underlying coupled-cluster approximation, and where the present Jastrow form fails. To answer these questions, we analyze basis-set trends, a simple $\Delta$MP2 correction, interaction-class-resolved errors, and the redistribution of reference and correlation contributions to the interaction energy.

The paper is organized as follows: Section~II introduces the transcorrelated Hamiltonian, the Jastrow optimization strategy, and the correction schemes used in this work. Section~III presents the computational details and Section~IV the A24 benchmark results. Section~V summarizes the conclusions and discusses future applications to larger noncovalent systems.

\section{Theory}
\subsection{Transcorrelation}

The transcorrelated method introduces explicit electron correlation by writing the many-electron wavefunction as a product of a Jastrow factor and a smoother auxiliary wavefunction,
\begin{equation}
\Psi = e^{\tau} \Phi ,
\label{eq:tc_ansatz}
\end{equation}
where $\tau$ is a real-space correlation function.  In the present study, we consider a Jastrow function which is the sum of one-electron and two-electron terms:
\begin{equation}
\tau = \tau_1 + \tau_2 =  \sum_i \chi(\mathbf{r}_i) + 
\sum_{i<j} u(\mathbf{r}_i,\mathbf{r}_j)
\equiv
\sum_i \chi_i + \sum_{i<j} u_{ij}.
\label{eq:tau_pair}
\end{equation}
Note that this treatment is slightly different to our earlier treatments of the TC method \cite{luo2018fciqmc, schraivogel2021tccc, christlmaier2023xtc}, in which the $\chi$ terms were combined into the $u$ terms. In an exact treatment of the three-body TC interactions, these two methods are equivalent. However, in an approximate treatment (the xTC approximation, which we detail shortly) the {\em separate} form of the Jastrow is more accurate, and for weak non-covalent interactions this turns out to be consequential. The resulting equations are slightly more involved, and are explicitly derived below and in the supplementary material. 

Substitution of Eq.~\eqref{eq:tc_ansatz} into the Schr\"odinger equation leads to a similarity-transformed Hamiltonian,
\begin{equation}
\hat{H}_{\mathrm{TC}}
=
e^{-\tau} \hat{H} e^{\tau},
\label{eq:tc_similarity}
\end{equation}
which acts on the auxiliary wavefunction $\Phi$. The transformation is non-unitary meaning that the transcorrelated Hamiltonian is, in general, non-Hermitian.

For the nonrelativistic electronic Hamiltonian, the Baker--Campbell--Hausdorff expansion terminates exactly,
\begin{equation}
\hat{H}_{\mathrm{TC}}
=
\hat{H}
+
[\hat{H},\tau]
+
\frac{1}{2}
[[\hat{H},\tau],\tau],
\label{eq:tc_bch}
\end{equation}
because $\tau$ is a multiplicative operator in real space. Equivalently, using
\begin{equation}
\hat{H}
=
-\frac{1}{2}\sum_i \nabla_i^2
+
\hat{V},
\end{equation}
with all potential-energy terms collected in $\hat{V}$, one obtains
\begin{equation}
\hat{H}_{\mathrm{TC}}
=
\hat{H}
-
\sum_i
\left[
\nabla_i \tau \cdot \nabla_i
+
\frac{1}{2}\nabla_i^2 \tau
+
\frac{1}{2}
\left(\nabla_i \tau\right)^2
\right].
\label{eq:tc_real_space_compact}
\end{equation}
For the separated one- and two-electron Jastrow form of Eq.~\eqref{eq:tau_pair},
Eq.~\eqref{eq:tc_real_space_compact} can be separated into one-, two-, and three-electron contributions,
\begin{equation}
\hat{H}_{\mathrm{TC}}
=
\hat{H}
+
\hat{M}
+
\hat{K}
+
\hat{L},
\label{eq:tc_hkl}
\end{equation}
where the one-electron contribution arising from $\chi_i$ is
\begin{equation}
\hat{M}
=
-\sum_i
\left[
\nabla_i\chi_i\cdot\nabla_i
+
\frac{1}{2}\nabla_i^2\chi_i
+
\frac{1}{2}\left|\nabla_i\chi_i\right|^2
\right],
\label{eq:tc_m_operator}
\end{equation}
and the two-electron contribution is
\begin{align}
\hat{K}
=
-\sum_{i<j}
\Bigg[
&
\nabla_i u_{ij}\cdot\nabla_i
+
\nabla_j u_{ij}\cdot\nabla_j
+
\frac{1}{2}
\left(
\nabla_i^2 u_{ij}
+
\nabla_j^2 u_{ij}
\right)
\nonumber
\\
&
+
\frac{1}{2}
\left(
\left|\nabla_i u_{ij}\right|^2
+
\left|\nabla_j u_{ij}\right|^2
\right)
\nonumber
\\
&
+
\nabla_i\chi_i\cdot\nabla_i u_{ij}
+
\nabla_j\chi_j\cdot\nabla_j u_{ij}
\Bigg].
\label{eq:tc_k_operator}
\end{align}
The last two terms in Eq.~\eqref{eq:tc_k_operator} arise from the cross terms between the one- and two-electron parts of the Jastrow factor in
$\left(\nabla_i\tau\right)^2$.
The remaining three-electron contribution is
\begin{equation}
\hat{L}
=
-
\sum_i
\sum_{\substack{j<k\\ j,k\ne i}}
\nabla_i u_{ij}\cdot\nabla_i u_{ik}.
\label{eq:tc_l_operator}
\end{equation}
Thus, $\hat{M}$ is an effective one-electron operator, $\hat{K}$ contains effective two-electron terms, and $\hat{L}$ is an effective three-electron operator. The appearance of $\hat{L}$ is a central practical difficulty of the transcorrelated method: an explicit treatment of the resulting three-body matrix elements greatly increases memory requirements and complicates the use of standard many-body electronic-structure algorithms. In the separated representation, all $\chi$-dependent contributions are retained explicitly in the one- and two-electron operators $\hat{M}$ and $\hat{K}$.

We avoid the explicit treatment of the three-body operator by employing the xTC approximation introduced in Ref.~\cite{christlmaier2023xtc}. The main idea of xTC is to normal order the three-body contribution with respect to a reference wavefunction and to retain its scalar, one-body, and two-body contractions, while discarding the remaining explicit normal-ordered three-body part. In this way, the dominant mean-field effects of $\hat{L}$ are folded into an effective Hamiltonian containing no operators higher than two-body rank.

The resulting xTC Hamiltonian can be written schematically as
\begin{equation}
\hat{H}_{\mathrm{xTC}}
=
\hat{H}
+
\hat{M}
+
\hat{K}
+
\hat{L}^{[0]}
+
\hat{L}^{[1]}
+
\hat{L}^{[2]},
\label{eq:xtc_hamiltonian_operator}
\end{equation}
where $\hat{L}^{[0]}$, $\hat{L}^{[1]}$, and $\hat{L}^{[2]}$ denote the zero-, one-, and two-body normal-ordered contractions of the three-body operator $\hat{L}$. Equivalently, in second quantization,
\begin{equation}
\hat{H}_{\mathrm{xTC}}
=
E_{\mathrm{xTC}}^{[0]}
+
\sum_{pq}
h^{\mathrm{xTC}}_{pq}
a_p^\dagger a_q
+
\frac{1}{4}
\sum_{pqrs}
v^{\mathrm{xTC}}_{pqrs}
a_p^\dagger a_q^\dagger a_s a_r .
\label{eq:xtc_second_quantized}
\end{equation}
Expressions for $E_{\mathrm{xTC}}^{[0]}, h^{\mathrm{xTC}}_{pq}$ and $v^{\mathrm{xTC}}_{pqrs}$ are provided in the Supplementary Information. Note that the xTC Hamiltonian has the same operator rank as the conventional electronic Hamiltonian and can be used in standard coupled-cluster calculations with essentially the same number of one- and two-body matrix elements. The main formal difference is that the transformed integrals are generally non-Hermitian, reflecting the non-unitary nature of the Jastrow similarity transformation.

\subsection{Jastrow factor and optimization strategy}
\label{sec:jastrow_optimization}
We use the flexible Jastrow form introduced by Drummond, Towler, and Needs (DTN), which has been widely used in quantum Monte Carlo and transcorrelated calculations \cite{drummond2004jastrow,haupt2023optimizing}. The Jastrow exponent is written as a sum of electron--electron, electron--nucleus, and electron--electron--nucleus terms,
\begin{equation}
\tau
=
\sum_{i<j} u(r_{ij})
+
\sum_{iI} \chi_I(r_{iI})
+
\sum_{i<j,I} f_I(r_{ij},r_{iI},r_{jI}),
\label{eq:dtn_jastrow}
\end{equation}
where $i,j$ label electrons and $I$ labels nuclei. The electron--electron term describes pair correlation and imposes the electron--electron cusp, the electron--nucleus term allows the Jastrow factor to adapt locally around each atom, and the electron--electron--nucleus term introduces environment-dependent pair correlation around the nuclei.

The individual terms are expanded in polynomial bases multiplied by finite cutoff functions. For the electron--electron term we use the form
\begin{equation}
u(r_{ij})
=
\left(1-\frac{r_{ij}}{L_{\mathrm{ee}}}\right)^{C}
\sum_{n=1}^{N_u} c_n r_{ij}^{\,n-1},
\qquad
r_{ij}<L_{\mathrm{ee}},
\label{eq:ee_jastrow}
\end{equation}
and $u(r_{ij})=0$ for $r_{ij}\geq L_{\mathrm{ee}}$. With this indexing convention, the electron--electron cusp condition is enforced by constraining the constant coefficient according to
\begin{equation}
c_1
=
\frac{L_{\mathrm{ee}}}{C}
\left(
c_2-\frac{1}{2}
\right),
\label{eq:ee_cusp_constraint}
\end{equation}
which ensures $u'(0)=1/2$ for antiparallel spin electron pairs, as required by the Kato cusp condition \cite{kato1957eigenfunctions}. The same idea can be generalized to spin-dependent cusp constraints, but in the present work we use the spin-averaged form.

The electron--nucleus term is written as
\begin{equation}
\chi_I(r_{iI})
=
\left(1-\frac{r_{iI}}{L_{\mathrm{en}}}\right)^{C}
\sum_{n=1}^{N_\chi} d_{nI} r_{iI}^{\,n-1},
\qquad
r_{iI}<L_{\mathrm{en}},
\label{eq:en_jastrow}
\end{equation}
and the electron--electron--nucleus term as
\begin{align}
f_I(r_{ij},r_{iI},r_{jI})
=
&
\left(1-\frac{r_{iI}}{L_{\mathrm{een}}}\right)^{C}
\left(1-\frac{r_{jI}}{L_{\mathrm{een}}}\right)^{C}
\nonumber
\\
&\times
\sum_{lmn=1}
g_{lmnI}
r_{ij}^{\,l-1}
r_{iI}^{\,m-1}
r_{jI}^{\,n-1},
\label{eq:een_jastrow}
\end{align}
with $f_I=0$ whenever either electron lies outside the cutoff sphere around nucleus $I$. In all calculations, the cutoff lengths are treated as fixed hyperparameters rather than variational parameters. Unless stated otherwise, we use
\begin{equation}
L_{\mathrm{ee}}=4.5\,a_0,
\qquad
L_{\mathrm{en}}=1.0\,a_0,
\qquad
L_{\mathrm{een}}=2.0\,a_0.
\label{eq:jastrow_cutoffs}
\end{equation}
This choice corresponds to the standard short-range Jastrow parametrization used in previous applications to covalent systems. A central question of the present work is whether the same local parametrization remains balanced for noncovalent interaction energies, where the energy scale is much smaller and the dimer--monomer cancellation is more delicate.

The Jastrow parameters are optimized by minimizing the variance of the transcorrelated reference energy \cite{haupt2023optimizing,filip2025deterministic}. For a reference determinant $\Phi_0$, the transcorrelated reference energy is
\begin{equation}
E_{\mathrm{ref}}
=
\langle \Phi_0 | \hat{H}_{\mathrm{TC}} | \Phi_0 \rangle .
\label{eq:tc_ref_energy}
\end{equation}
The corresponding reference variance is
\begin{equation}
\sigma_{\mathrm{ref}}^2
=
\langle \Phi_0 |
\left|
\hat{H}_{\mathrm{TC}}-E_{\mathrm{ref}}
\right|^2
| \Phi_0 \rangle .
\label{eq:ref_variance}
\end{equation}
By inserting a resolution of the identity in the determinant basis, this can be written as
\begin{equation}
\sigma_{\mathrm{ref}}^2
=
\sum_{I\ne 0}
\left|
\langle \Phi_I |
\hat{H}_{\mathrm{TC}}
| \Phi_0 \rangle
\right|^2 ,
\label{eq:ref_variance_determinants}
\end{equation}
where the sum runs over determinants coupled to the Hartree--Fock reference by the transcorrelated Hamiltonian. Minimizing Eq.~(\ref{eq:ref_variance_determinants}) suppresses the coupling of the reference determinant to the determinant space connected by the transcorrelated Hamiltonian. In the xTC Hamiltonian, this primarily targets the reference-to-double-excitation couplings that enter the subsequent coupled-cluster treatment. In this sense, the Jastrow transformation compactifies the many-body wavefunction seen by the subsequent post-Hartree--Fock method, such as coupled-cluster theory.

As such, the role of the Jastrow factor is twofold. First, the fixed electron--electron cusp condition removes the Coulomb singularity from the auxiliary wavefunction and accelerates convergence with respect to the one-particle basis set. Second, the optimization of the remaining Jastrow parameters aims to make the right eigenfunction of the transcorrelated Hamiltonian more compact in determinant space. This second effect distinguishes the transcorrelated approach from conventional F12 methods, which primarily accelerate the basis-set convergence of a chosen many-body approximation without changing the complete-basis-set limit of that approximation \cite{kong2012f12}. In contrast, a truncated post-Hartree--Fock method applied to a transcorrelated Hamiltonian will not, in general, converge to the same complete-basis-set limit as the corresponding method applied to the bare Coulomb Hamiltonian.

For noncovalent interaction energies, the optimization protocol itself becomes an important part of the method. A straightforward strategy would be to optimize separate Jastrow factors for the dimer and for each monomer. However, this is problematic for two reasons. First, if the Jastrow parameters are obtained from variational Monte Carlo sampling, the stochastic uncertainty in the optimized parameters and in the resulting reference energies can be comparable to the small energy scale of noncovalent binding. Reducing this noise would require a very large number of configurations and quickly becomes impractical for systematic benchmark studies. Second, separate optimizations can introduce a monomer--dimer imbalance. With a single global electron--electron term, the dimer Jastrow has to describe intramonomer correlation in both fragments as well as interfragment correlation. The monomer Jastrow, by contrast, only has to describe a simpler intramonomer environment with the same number of parameters. Separate optimization can therefore over-adapt the monomer Jastrow relative to the dimer Jastrow and bias the interaction energy.

To reduce this imbalance, we optimize the Jastrow parameters only for the dimer and reuse the same parameters for the corresponding monomers. This shared-parameter protocol is designed to improve error cancellation between the dimer and monomer calculations. Its success relies on the transferability of the dimer Jastrow to the monomer fragments. We promote this transferability by keeping the nucleus-centered terms local through short electron--nucleus and electron--electron--nucleus cutoffs. As a result, the local environment within each monomer is described by the same global electron--electron term together with local nucleus-centered terms, while the global electron--electron term remains available to describe interfragment correlation in the dimer. In this way, the dimer and monomer calculations use a consistent Jastrow description without requiring separate stochastic optimizations for each fragment.

\subsection{Correction schemes}

We consider two additive correction schemes. Both corrections follow the general spirit of focal-point or composite energy schemes, in which different energy contributions are evaluated at different levels of cost and basis-set quality \cite{csaszar1998fpa,rezac2015extensions}. The first scheme is the reference-corrected xTC scheme, RC-xTC, introduced in Ref.~\cite{hauskrecht2026referencecorrection}. This correction was originally motivated by the observation that accurate energy differences in low-quality basis sets, in particular atomization energies, can be limited by the convergence of the reference contribution rather than by the correlation energy. Let $S$ denote the target small basis and $L$ a larger basis. In this work, \(S\) denotes either AVDZ or AVTZ, while \(L=\) AVQZ unless stated otherwise. The Jastrow factor $\tau_L$ is optimized in the large basis and then reused in the small-basis calculation. The reference-corrected energy is defined as
\begin{align}
E^{\mathrm{RC\mbox{-}xTC}}(\tau_L;S,L)
=
E_{\mathrm{ref}}^{\mathrm{xTC}}(\tau_L;L)
+
E_{\mathrm{corr}}^{\mathrm{xTC}}(\tau_L;S).
\label{eq:rc_xtc_energy}
\end{align}
Thus, the correlated calculation is still performed only in the small basis, while the reference contribution is imported from the larger basis. For interaction energies, Eq.~\eqref{eq:rc_xtc_energy} is applied consistently to the dimer and to the corresponding monomers. 

The second correction is a simple $\Delta$MP2 correction. It is designed to estimate the leading correlation-basis effect beyond AVDZ without requiring a coupled-cluster calculation in the larger basis. Starting from the RC-xTC energy in the small basis $S$, we add the change in the xTC-MP2 correlation energy between $S$ and an intermediate basis $T$,
\begin{equation}
\begin{split}
E^{\mathrm{RC\mbox{-}xTC}+\Delta\mathrm{MP2}}(S,T,L)
&=
E^{\mathrm{RC\mbox{-}xTC}}(S,L)
\\
\quad+
&\left[
E_{\mathrm{corr}}^{\mathrm{xTC-MP2}}(T)
-
E_{\mathrm{corr}}^{\mathrm{xTC-MP2}}(S)
\right].
\end{split}
\label{eq:delta_mp2_total}
\end{equation}
In the present work, $S=\mathrm{AVDZ}$, $T=\mathrm{AVTZ}$, and $L=\mathrm{AVQZ}$.
We test this correction because it is a practical route toward larger noncovalent complexes. For systems beyond A24, coupled-cluster calculations in a triple-zeta basis may become prohibitively expensive, whereas MP2 remains considerably cheaper and requires only two-electron integral information with at most two virtual orbital indices. In this way, the $\Delta$MP2 correction provides a low-cost estimate of the remaining correlation-basis improvement on top of the small-basis RC-xTC result \cite{moller1934note}.

\section{Computational Details}
In the following, we benchmark interaction energies for the A24 dataset against nonrelativistic electronic reference values. The reference interaction energy is constructed from the  CCSD(T)/CBS component reported for the A24 dataset \cite{rezac2013goldstandard}, supplemented by the nonrelativistic higher-level corrections,
\begin{equation}
E_{\mathrm{int}}^{\mathrm{ref}}
=
E_{\mathrm{int}}^{\mathrm{CCSD(T)/CBS}}
+
\Delta E_{\mathrm{int}}^{\mathrm{core}}
+
\Delta E_{\mathrm{int}}^{\mathrm{CCSDT(Q)}} ,
\label{eq:a24_reference_energy}
\end{equation}
where $\Delta E_{\mathrm{int}}^{\mathrm{core}}$ accounts for the difference between frozen-core and all-electron correlation, and $\Delta E_{\mathrm{int}}^{\mathrm{CCSDT(Q)}}$ estimates the post-CCSD(T) contribution. The CCSD(T)/CBS component was obtained from large augmented correlation-consistent basis sets, while the core-correlation correction was evaluated from all-electron and frozen-core CCSD(T) calculations with core-valence basis sets. The post-CCSD(T) correction was taken from the highest-level CCSDT(Q) data available for each A24 complex, using the basis sets reported in the original A24 benchmark studies \cite{rezac2013goldstandard,rezac2015extensions}. Relativistic corrections are not included in Eq.~\eqref{eq:a24_reference_energy}, since the present calculations are compared on the nonrelativistic electronic energy scale \cite{rezac2013goldstandard,rezac2015extensions}.

All calculations in this work are performed as all-electron (AE) calculations in the aVDZ and aVTZ basis sets. For the reference-corrected transcorrelated calculations, the reference contribution is evaluated in the larger aVQZ basis, as described in Eq.~\eqref{eq:rc_xtc_energy}. The Jastrow factor is also optimized in this basis set. For all dimers and monomers, full counterpoise correction is applied in order to correct for basis set superposition error in the interaction energies \cite{boys1970counterpoise,marshall2011basis}.
The AE calculations were used here to allow a direct comparison with the nonrelativistic all-electron reference energies. 

We compare three coupled-cluster levels, CCSD, distinguishable-cluster singles and doubles (DCSD), and CCSD(T), in three frameworks. The first is the conventional canonical formulation with the untransformed Coulomb Hamiltonian, which we refer to as the plain method. The second is the explicitly correlated F12 framework. All F12 calculations reported in this work use the F12b approximation together with the fixed-amplitude ansatz
\cite{sirianni2017F12NCI,knizia2009f12,kesharwani2018S66F12,kats2015}. The third is the reference-corrected transcorrelated approach, denoted RC-xTC. In the following, this approach is referred to simply as xTC for brevity. This comparison is designed to separate two effects. First, we assess how efficiently F12 and xTC accelerate the basis-set convergence of the interaction energy. Second, we examine whether the transcorrelated Hamiltonian changes the performance of the underlying post-Hartree--Fock approximation itself. The latter point is particularly important because, unlike F12 corrections, the transcorrelated transformation changes the effective Hamiltonian on which the approximate coupled-cluster method acts.

The $\Delta$MP2 correction defined in Eq.~\eqref{eq:delta_mp2_total} is also applied to the plain and F12 calculations. For the plain calculations, the correction is constructed from the difference of conventional MP2 correlation energies between AVTZ and AVDZ. For the F12 calculations, the analogous correction is obtained from the corresponding MP2-F12 correlation energies. Thus, the same additive correction strategy is used for all three frameworks, but with the MP2-level correlation energy chosen consistently with the underlying Hamiltonian or explicitly correlated treatment.

The plain and F12 coupled-cluster calculations were performed with the MOLPRO quantum chemistry package \cite{werner2012molpro}. The Jastrow parameters were optimized at the variational Monte Carlo level using the CASINO code \cite{needs2020casino}. The transcorrelated one- and two-electron integrals were generated with our in-house TCHInt program and subsequently passed through an FCIDUMP interface to the ElemCo.jl package for the xTC coupled-cluster calculations \cite{kats2024orbital,elemcojl}.


\section{Results}
\subsection{Basis-set and method dependence}

\begin{figure*}[tbp]
\centering
\includegraphics[width=\textwidth]{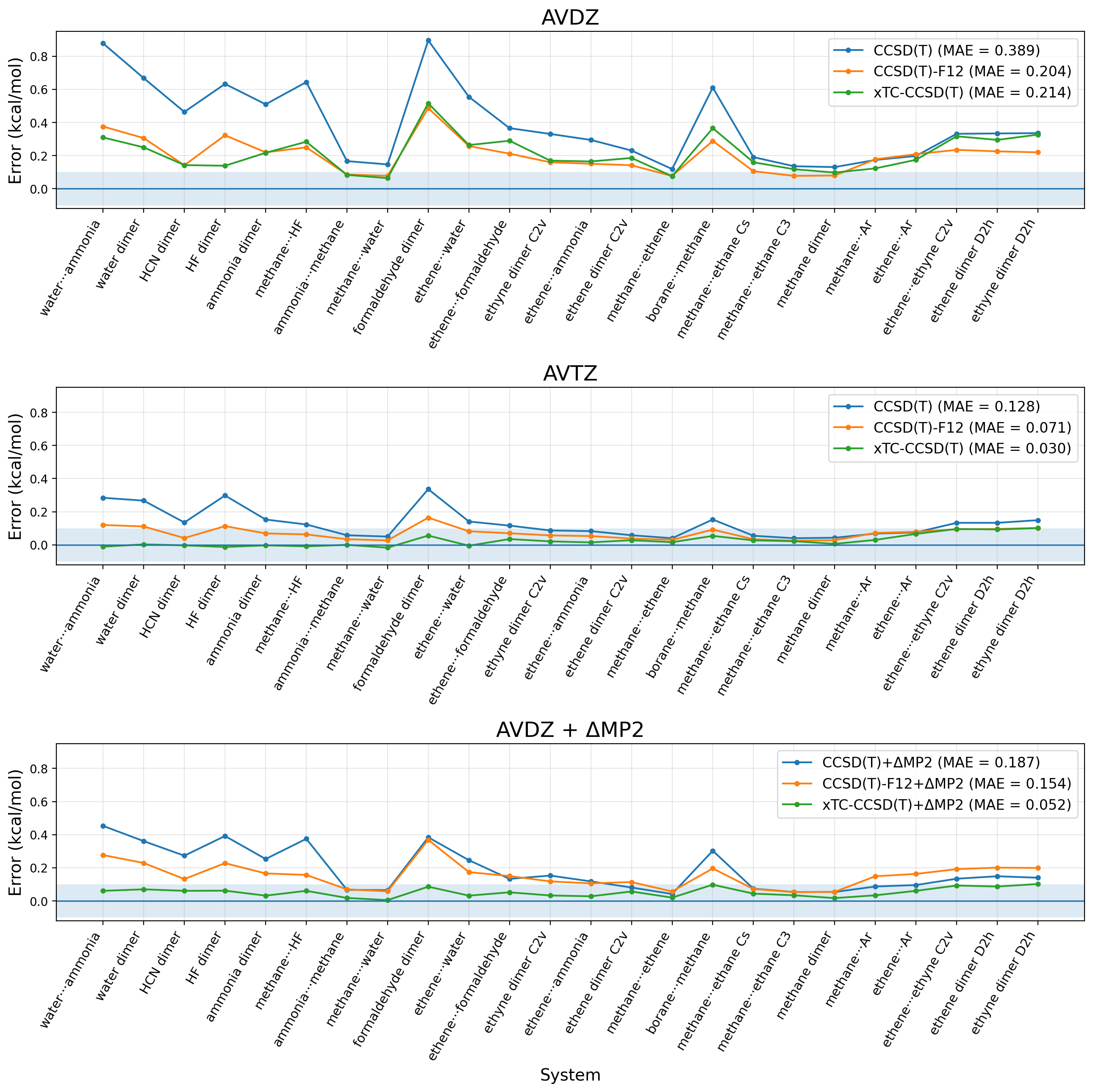}
\caption{
System-resolved absolute interaction-energy errors for the A24 dataset at the CCSD(T) level. Results are shown for AVDZ, AVTZ, and AVDZ supplemented by the $\Delta$MP2 correction. Plain CCSD(T), CCSD(T)-F12, and xTC-CCSD(T) are compared against the nonrelativistic electronic A24 reference values. The shaded region indicates a stringent $\pm 0.1$ kcal mol$^{-1}$ error window, a commonly used high-accuracy target for benchmark noncovalent interaction energies \cite{karton2023benchmark,marshall2011basis}. In AVDZ, CCSD(T)-F12 and xTC provide a comparable improvement over the plain calculations, although most systems do not fall within the shaded region. In AVTZ, xTC yields consistently smaller errors than CCSD(T)-F12 for most systems, leading to a substantially lower overall MAE. The same trend is observed for the AVDZ+$\Delta$MP2 results, demonstrating that the $\Delta$MP2 correction preserves the improved performance of xTC while providing a computationally efficient route toward triple-$\zeta$ accuracy.
}
\label{fig:a24_ccsdt_systemwise}
\end{figure*}

\begin{figure*}[tbp]
\centering
\includegraphics[width=\textwidth]{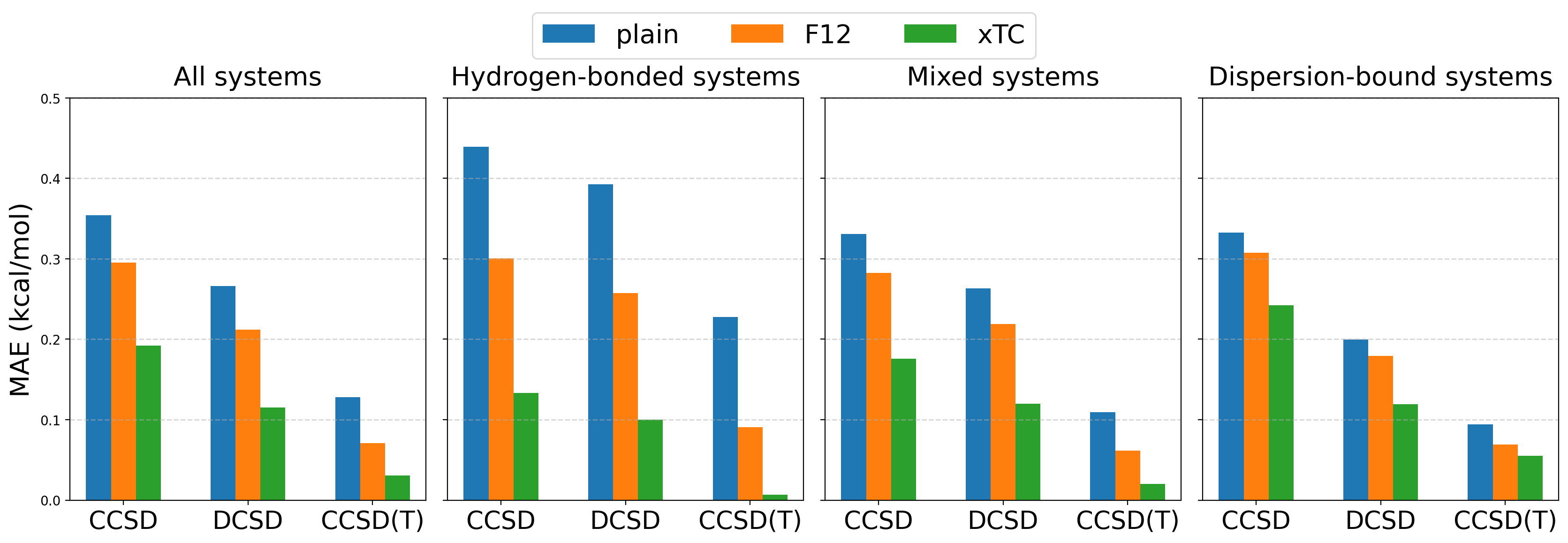}
\caption{
Mean absolute errors for the A24 interaction energies in the AVTZ basis, resolved by coupled-cluster level and interaction class. The MAEs of CCSD, DCSD, and CCSD(T) are shown for plain, F12, and xTC calculations. The four panels correspond to the full A24 set, hydrogen-bonded systems, mixed electrostatic/dispersion systems, and dispersion-dominated systems. In AVTZ, xTC gives the lowest MAEs for the full A24 set at all three coupled-cluster levels. The improvement over F12 is most pronounced for the hydrogen-bonded and mixed subsets, especially at the CCSD and DCSD levels. For the dispersion-dominated systems, the xTC advantage is smaller but remains visible across all three correlation levels, indicating that the transcorrelated Hamiltonian improves the effective many-body problem throughout the AVTZ benchmark.
}
\label{fig:a24_method_dependence_avtz}
\end{figure*}

We first consider the system-resolved CCSD(T) errors shown in Fig.~\ref{fig:a24_ccsdt_systemwise}. The shaded region indicates a stringent $\pm 0.1$ kcal mol$^{-1}$ error window, a commonly used high-accuracy target for benchmark noncovalent interaction energies \cite{karton2023benchmark,marshall2011basis}. In the double-zeta basis, plain CCSD(T) exhibits substantial errors for most of the A24 set, with a MAE of $0.389$ kcal mol$^{-1}$ (Table~\ref{tab:a24_mae_all}). Both F12 and xTC substantially reduce these errors, yielding AVDZ MAEs of $0.204$ and $0.214$ kcal mol$^{-1}$, respectively. Thus, in contrast to the smaller errors obtained in larger bases, the double-$\zeta$ results show that both explicitly correlated approaches provide a comparable improvement over the plain calculation, but neither is sufficient to bring most systems below the stringent $\pm 0.1$ kcal mol$^{-1}$ error window. AVDZ therefore remains a demanding regime for uniformly high-accuracy A24 interaction energies.


The situation changes more clearly in AVTZ. Here, both explicitly correlated approaches reduce the plain CCSD(T) error substantially, but xTC-CCSD(T) gives the lower overall MAE. The MAE decreases from $0.128$ kcal mol$^{-1}$ for plain CCSD(T) to $0.071$ kcal mol$^{-1}$ for CCSD(T)-F12 and to $0.030$ kcal mol$^{-1}$ for xTC-CCSD(T). Thus, in the triple-$\zeta$ basis, xTC not only becomes competitive with F12, but yields a visibly more balanced system-resolved error pattern. In particular, the xTC-CCSD(T) values for the first eight systems are nearly indistinguishable from the reference values on the scale of Fig.~\ref{fig:a24_ccsdt_systemwise}, indicating very strong performance for hydrogen-bonded and mixed electrostatic complexes.


A similar trend is observed when the AVDZ results are supplemented by the $\Delta$MP2 correction. The correction lowers the xTC-CCSD(T) MAE from $0.214$ to $0.052$ kcal mol$^{-1}$, bringing the small-basis transcorrelated result close to the AVTZ accuracy. The improvement is less pronounced for CCSD(T)-F12, for which the AVDZ+$\Delta$MP2 MAE is $0.154$ kcal mol$^{-1}$. Thus, within the present correction scheme, $\Delta$MP2 is particularly effective for the xTC interaction energies.

While Fig.~\ref{fig:a24_ccsdt_systemwise} focuses on the basis-set dependence at the CCSD(T) level, Fig.~\ref{fig:a24_method_dependence_avtz} isolates the method dependence in the AVTZ basis. The figure shows the MAEs of CCSD, DCSD, and CCSD(T) for the plain, F12, and xTC approaches. To reveal interaction-class-dependent trends, the statistics are shown not only for the full A24 set, but also separately for the hydrogen-bonded, mixed electrostatic/dispersion, and dispersion-dominated subsets. The corresponding full-set MAEs are also summarized in Table~\ref{tab:a24_mae_all}. For the full A24 set, xTC gives the lowest MAEs at all three coupled-cluster levels, reducing the errors relative to both the plain and F12 calculations. This advantage is most pronounced for the lower-level CCSD and DCSD approximations, but it remains clearly visible at the CCSD(T) level as well. For the hydrogen-bonded systems, xTC is particularly effective and gives by far the smallest MAEs across all three correlation levels. A similarly favorable trend is observed for the mixed systems. Even for the dispersion-dominated subset, where all methods perform more similarly, xTC still yields the lowest MAEs at the CCSD, DCSD, and CCSD(T) levels.

\begin{table}[tbp]
\centering
\scriptsize
\setlength{\tabcolsep}{4pt}
\renewcommand{\arraystretch}{1.08}
\caption{Mean absolute errors (MAEs, kcal mol$^{-1}$) in AVTZ for all A24 systems and all interaction classes.}
\label{tab:a24_mae_all}
\begin{tabular}{@{}llrrr@{}}
\toprule
Interaction class & Variant & CCSD & DCSD & CCSD(T) \\
\midrule
All systems & plain & 0.354 & 0.266 & 0.128 \\
& F12 & 0.296 & 0.212 & 0.071 \\
& xTC & 0.192 & 0.115 & 0.030 \\
\addlinespace[0.15em]
Hydrogen-bonded & plain & 0.439 & 0.393 & 0.227 \\
& F12 &  0.300 & 0.258 & 0.091 \\
& xTC & 0.133 & 0.100 & 0.007 \\
\addlinespace[0.15em]
Mixed & plain & 0.331 & 0.263 & 0.109 \\
& F12 & 0.282 & 0.219 & 0.062 \\
& xTC & 0.176 & 0.120 & 0.020 \\
\addlinespace[0.15em]
Dispersion & plain & 0.333 & 0.199 & 0.094 \\
& F12 & 0.308 & 0.179 & 0.069 \\
& xTC & 0.242 & 0.119 & 0.055 \\
\bottomrule
\end{tabular}
\end{table}


These trends support the method-level improvement discussed in Sec.~\ref{sec:jastrow_optimization}: the transcorrelated Hamiltonian improves the effective many-body problem, and this effect is most pronounced when the underlying coupled-cluster approximation is less complete. The largest gains are therefore observed for CCSD and DCSD. Importantly, however, the advantage does not disappear at the CCSD(T) level in AVTZ, where xTC still yields lower MAEs than the corresponding F12 calculations. This indicates that the transformed Hamiltonian provides a systematic improvement beyond pure basis-set acceleration for the present benchmark.

\begin{figure*}[tbp]
\centering
\includegraphics[width=\textwidth]{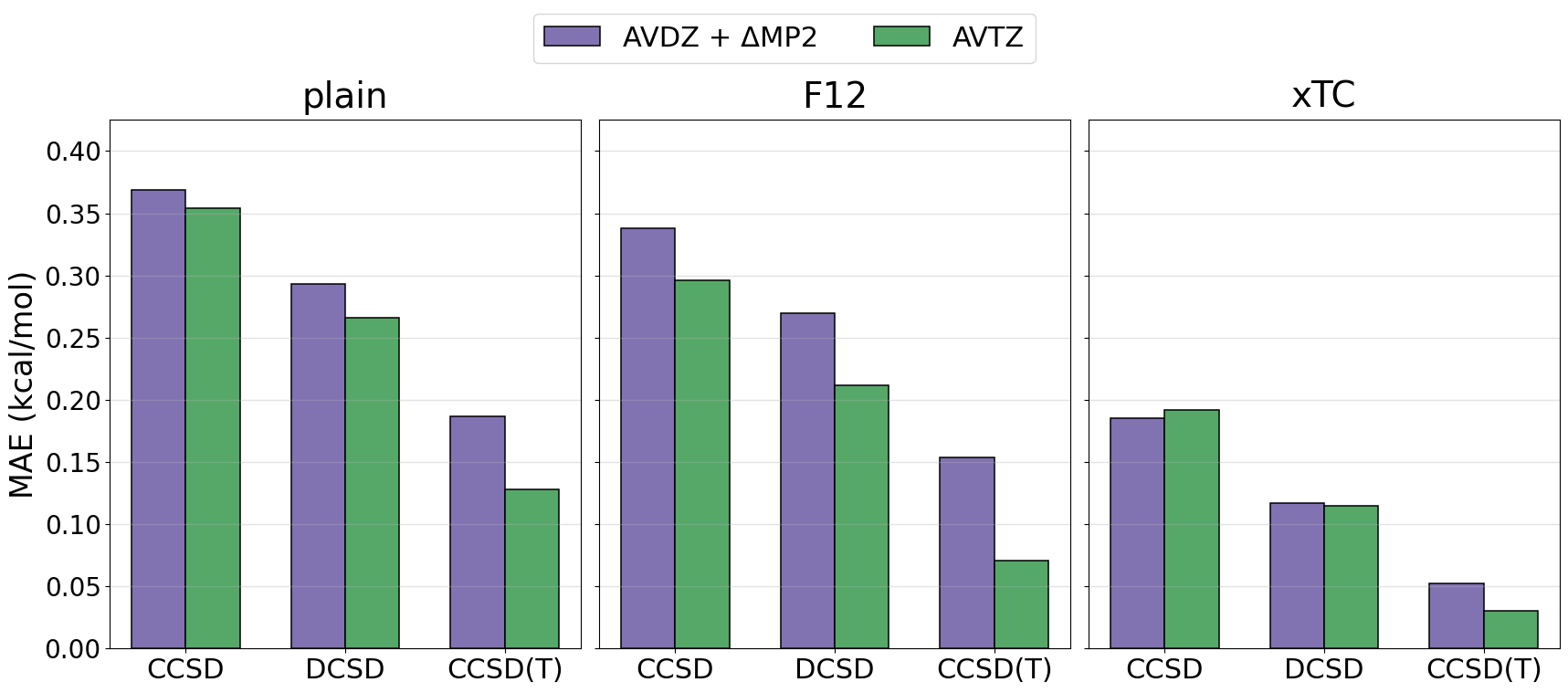}
\caption{
Comparison of AVDZ+$\Delta$MP2 and AVTZ mean absolute errors for the A24 interaction energies. The three panels correspond to plain, F12, and xTC calculations, and each panel shows the MAEs at the CCSD, DCSD, and CCSD(T) levels. The AVDZ+$\Delta$MP2 protocol reproduces the AVTZ accuracy particularly well for xTC at the CCSD and DCSD levels, where the two sets of MAEs are nearly identical. A somewhat larger difference remains at the CCSD(T) level, although the absolute xTC+$\Delta$MP2 error remains small. For the plain and F12 calculations, AVDZ+$\Delta$MP2 follows the AVTZ trend qualitatively but gives larger deviations, especially at the CCSD(T) level. Overall, the comparison shows that the $\Delta$MP2 correction is especially effective in combination with xTC and provides a low-cost route toward triple-$\zeta$ accuracy for the transcorrelated calculations.
}
\label{fig:a24_delta_mp2_vs_avtz}
\end{figure*}



We next test whether the cheaper AVDZ+$\Delta$MP2 protocol can reproduce the AVTZ trends. Figure~\ref{fig:a24_delta_mp2_vs_avtz} compares the MAEs obtained with AVDZ+$\Delta$MP2 to the corresponding AVTZ values for plain, F12, and xTC calculations at the CCSD, DCSD, and CCSD(T) levels. The comparison shows that the effectiveness of the $\Delta$MP2 correction is strongly method dependent. For the plain calculations, AVDZ+$\Delta$MP2 captures part of the improvement associated with increasing the basis to AVTZ, but the agreement with the AVTZ results is not uniform across correlation levels. For F12, the same correction is less successful: the AVDZ+$\Delta$MP2 MAEs deviate more strongly from the corresponding AVTZ values at all three coupled-cluster levels, with the largest discrepancy occurring for CCSD(T).

The most consistent behavior is obtained for xTC. Here, the AVDZ+$\Delta$MP2 and AVTZ MAEs are nearly identical at the CCSD and DCSD levels, and the CCSD(T) value remains low despite a somewhat larger difference from the AVTZ result. This indicates that, for the present dataset, the $\Delta$MP2 correction is particularly effective in combination with the transcorrelated Hamiltonian. The correction, therefore, provides a practical low-cost route to approach AVTZ-quality interaction energies for xTC.

This result is of practical importance beyond the present A24 benchmark. For larger noncovalent systems, full coupled-cluster calculations in a triple-zeta basis can become difficult, while MP2 calculations remain considerably more feasible. The $\Delta$MP2 correction therefore provides a simple way to improve the accuracy of the correlation energies without requiring the full post-Hartree--Fock calculation in the larger basis.

\subsection{Reference--correlation redistribution}
\begin{figure*}[tbp]
\centering
\includegraphics[width=\textwidth] {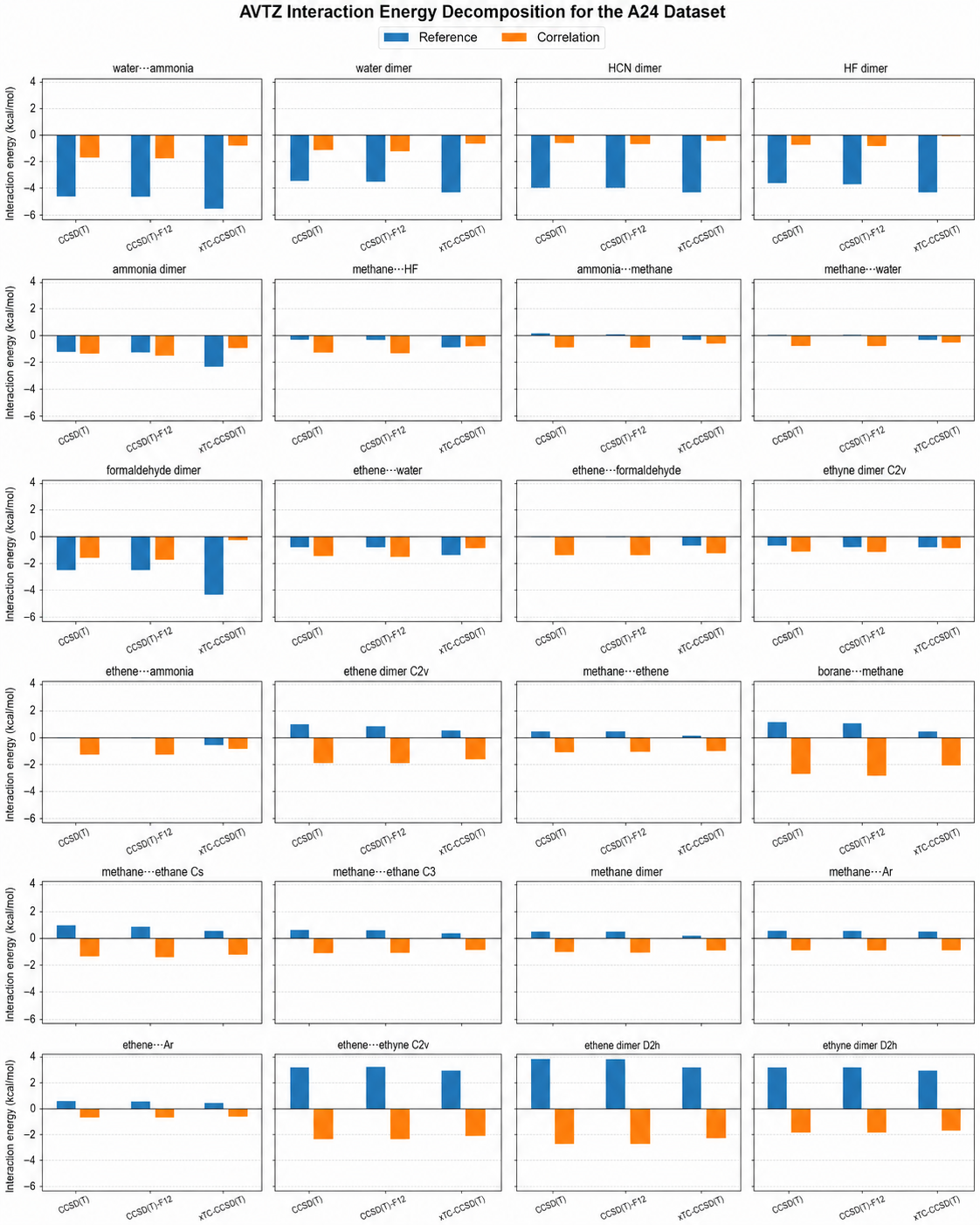}
\caption{
Reference and correlation contributions to the A24 interaction energies at the CCSD(T) level in AVTZ. Plain CCSD(T), CCSD(T)-F12, and xTC-CCSD(T) are compared. F12 leaves the reference contribution almost unchanged and mainly increases the attractive correlation contribution. In contrast, xTC shifts part of the interaction energy into the reference contribution, leading to a more attractive reference interaction and a less attractive remaining correlation contribution.
}
\label{fig:a24_ref_corr_decomp}
\end{figure*}

To understand how the transcorrelated Hamiltonian changes the description of the interaction energy, we decompose the total interaction energy schematically into reference and correlation contributions,
\begin{equation}
E_{\mathrm{int}}
=
E_{\mathrm{int}}^{\mathrm{ref}}
+
E_{\mathrm{int}}^{\mathrm{corr}},
\label{eq:eint_ref_corr_decomp}
\end{equation}
with
\begin{equation}
E_{\mathrm{int}}^{\lambda}
=
E_{AB}^{\lambda}
-
E_A^{\lambda}
-
E_B^{\lambda},
\qquad
\lambda \in \{\mathrm{ref},\mathrm{corr}\}.
\label{eq:eint_component}
\end{equation}
Figure~\ref{fig:a24_ref_corr_decomp} shows these two contributions, for every system of the A24 dataset, for the CCSD(T) calculations in AVTZ.

The comparison between plain CCSD(T) and CCSD(T)-F12 shows the expected behavior of an explicitly correlated basis-set correction. The reference contribution is almost unchanged by F12, apart from the small effect of the CABS singles correction, while the correlation contribution becomes more attractive. Thus, F12 primarily improves the correlation part of the interaction energy, consistent with its role as a basis-set accelerator for short-range pair correlation.

The transcorrelated Hamiltonian changes this decomposition in a qualitatively different way. Across the A24 set, xTC makes the reference contribution more attractive, or equivalently less repulsive, while the remaining correlation contribution becomes less attractive. In this sense, part of the interaction energy that appears as correlation energy for the bare Hamiltonian is shifted into the reference contribution of the transcorrelated Hamiltonian. However, the magnitude of this redistribution is strongly system dependent. For compact systems such as the HF dimer and the formaldehyde dimer, the xTC interaction energy is almost entirely contained in the reference contribution, with only a small correlation contribution. In contrast, for weakly bound dispersion systems such as ethene--Ar and methane--Ar, the redistribution is much smaller. Importantly, xTC does not qualitatively change the sign pattern of the decomposition: systems with attractive reference contributions remain reference-attractive, while dispersion-dominated systems remain characterized by a repulsive reference contribution and an attractive correlation contribution.

This system-resolved view also helps rationalize why the relative xTC advantage is smaller for the dispersion-dominated subset. In these complexes, the total interaction energy often results from a delicate cancellation between a repulsive reference term and an attractive correlation term. xTC reduces the magnitude of both contributions, but this redistribution does not necessarily improve the final cancellation in the total interaction energy. This is consistent with the class-resolved MAEs in Fig.~\ref{fig:a24_method_dependence_avtz}, where the relative advantage of xTC is smaller for the dispersion-dominated subset than for the hydrogen-bonded and mixed subsets.



\section{Conclusion and Outlook}

In this work, we have presented 
the first application of transcorrelated coupled-cluster theory to benchmark noncovalent interaction energies. The A24 dataset provides a stringent test for this purpose because it contains hydrogen-bonded, mixed electrostatic/dispersion, and dispersion-dominated complexes, and because the interaction energies are small differences between large monomer and dimer total energies. This makes the balance between dimer and monomer calculations particularly important.

A central methodological aspect of the present work is the Jastrow optimization strategy. Instead of optimizing independent Jastrow factors for dimers and monomers, we optimize the Jastrow parameters for the dimer and reuse them for the corresponding monomer calculations. This shared-parameter protocol is designed to reduce stochastic noise and avoid an imbalance between the flexibility of the monomer and dimer Jastrows. The use of local nucleus-centered terms together with a global electron--electron term allows the monomer environment to remain transferable, while retaining a description of interfragment correlation in the dimer.


The comparison with plain and F12 coupled-cluster calculations shows that both F12 and xTC substantially reduce the errors of the plain calculations already in the double-$\zeta$ basis. In AVDZ, the two approaches give comparable CCSD(T) accuracy, indicating that both treatments provide an effective small-basis correction, although neither yields uniformly high-accuracy interaction energies for the full A24 set. In AVTZ, however, the transcorrelated results become clearly more favorable. Across the CCSD, DCSD, and CCSD(T) levels, xTC yields lower MAEs than the corresponding F12 calculations, with the largest relative improvements observed for the lower-level CCSD and DCSD approximations. This indicates that TC does not only accelerate basis-set convergence, but also improves the effective many-body problem seen by the approximate coupled-cluster solver. Importantly, this method-level improvement remains visible at the CCSD(T) level, where xTC still outperforms F12 in the AVTZ basis.

The $\Delta$MP2 correction provides an important practical extension of this idea. By adding the difference between MP2 correlation energies in AVTZ and AVDZ to the small-basis xTC energy, one obtains results that approach the full AVTZ xTC calculations, especially at the CCSD and DCSD levels. At the CCSD(T) level, a somewhat larger difference to the AVTZ result remains, but the corrected small-basis error is still low. This is relevant for larger noncovalent systems, where full coupled-cluster calculations in triple-zeta bases may become computationally demanding, while MP2 calculations remain feasible. The xTC+$\Delta$MP2 protocol offers a practical route for applying transcorrelated methods beyond the small dimers of the A24 set.


The reference--correlation decomposition further shows that TC changes the structure of the interaction energy in a way that is qualitatively different from F12. While F12 primarily improves the correlation contribution and leaves the reference interaction almost unchanged, the transcorrelated Hamiltonian redistributes the interaction energy by shifting part of the attractive correlation contribution into the reference contribution. This redistribution illustrates that TC is not only a basis-set acceleration technique, but also modifies the effective many-body problem seen by the subsequent coupled-cluster treatment.

Overall, the present results suggest that transcorrelated coupled-cluster theory provides a promising alternative route to high-accuracy calculations of noncovalent interaction energies. Future work will focus on three directions. First, improved or more transferable Jastrow forms should be explored, especially for long-range dispersion-dominated interactions. Second, the basis set superposition error of xTC interaction energies should be investigated systematically. In the present work, all interaction energies were evaluated with full counterpoise correction, but the magnitude and distance dependence of the underlying BSSE, as well as its interplay with the reference correction, were not analyzed in detail. Third, we plan to apply the present methodology to larger noncovalent systems, in particular water hexamers and water clusters more generally. These systems provide a natural next step because they test not only pairwise noncovalent interactions, but also cooperative many-body hydrogen bonding and the transferability of the Jastrow factor across larger molecular assemblies.

\section*{Supplementary material}

The plain, F12b, and xTC interaction energies obtained with CCSD, DCSD, and CCSD(T) for all basis sets, together with the corresponding total energies and the xTC working equations for the separated Jastrow formalism, are provided in the Supplementary Material

\begin{acknowledgments}
The authors gratefully acknowledge the support of the Max Planck Society.
\end{acknowledgments}




\bibliographystyle{unsrtnat}
\bibliography{bib}

\end{document}